\documentclass[prb,showpacs,twocolumn,preprintnumbers,amsmath,amssymb]{revtex4}

\usepackage{epsfig}
\usepackage{graphicx}
\usepackage{dcolumn}
\usepackage{bm}

\newcommand{\vf}{v_{\rm F}}

\begin{document}

\title{Transport through evanescent waves in ballistic graphene quantum dots}

\author{M. I. Katsnelson$^1$}
\affiliation{$^1$ Institute for Molecules and Materials, Radboud
University Nijmegen, Heijendaalseweg 135, 6525 AJ, Nijmegen, The
Netherlands }
  \author{F. Guinea$^2$}
\affiliation{$^2$Instituto de Ciencia de Materiales de Madrid
(CSIC), Sor Juana In\'es de la Cruz 3, Madrid 28049, Spain}

\begin{abstract}
We study the transport through evanescent waves in graphene quantum
dots of different geometries. The transmission is suppressed when
the leads are attached to edges of the same majority sublattice.
Otherwise, the transmission depends exponentially on the distance
between leads in rectangular dots, and as a power law in circular
dots. The transmission through junctions where the transmitted and
reflected currents belong to the opposite valley as the incoming one
depends on details of the lattice structure at distances comparable
to the atomic spacing.
\end{abstract}
\pacs{73.20.-r; 73.21.La; 73.23.Ad}

\maketitle
\section{Introduction}
Since its synthesis and identification\cite{Netal04,Netal05b},
graphene has attracted a great deal of interest, because of its
unusual fundamental properties, and possible
applications\cite{GN07,KN07,NGPNG07}. Being purely planar material
(actually, graphene is the first example of truly two-dimensional
crystals) and demonstrating a very high charge carrier mobility
graphene is considered as a perspective base for a post-silicon
electronics.

The carriers in graphene are described by the massless Dirac
equation and possess a pseudospin degree of freedom (which is,
actually, a sublattice label) and chirality related with it. A
number of unusual properties follow from this, and, in particular,
the existence of localized states at the Dirac energy (that is, at
the neutrality point), and a new transport regime dominated by
evanescent waves, also at the Dirac energy\cite{K06,TTTRB06} (see
also the experimental studies in\cite{Metal07,Detal08}). Midgap
states with the energy close to zero (further we will count the
energy from the Dirac energy) were initially found at graphene edges
with a perfect termination in one of the two sublattices which
define the honeycomb lattice\cite{FWNK96} (zigzag edges), and later
they were generalized to other defects, such as cracks\cite{VLSG05},
vacancies\cite{PGLPN06}, generic surfaces with a majority of atoms
of one sublattice\cite{AB08}, and ripples\cite{GKV07}. These midgap
states have similar wave functions to the evanescent waves which
mediate the transport in clean graphene when there are no charge
carriers and the chemical potential coincides with the neutrality
point\cite{K06,TTTRB06}. The combination of evanescent waves and
localized states can even enhance the conductivity in graphene with
defects\cite{LVGC07,T07}. It can also be expected that midgap states
dominate the transport properties of graphene quantum
dots\cite{WSG08} which are the subject of intensive study
now\cite{Betal05,GN07,HOZK07,AZP07,Petal08}.

In the following, we extend the analysis the transport properties
of clean graphene with chemical potential equal to
zero\cite{K06,TTTRB06} to graphene quantum dots of different
geometries. We first present the model, and then we consider three
different cases, namely, rectangular dot, circular dot, and the
corner between two facets representative of broad classes of
quantum dots. In particular, we will demonstrate that for the case
of circular dot the conductance is very sensitive to magnetic
field which may give an insight for development a new type of
magnetic sensors.

The main conclusions can be found in the last section.
\section{The model}
We will consider ballistic quantum dots etched from a single layer
graphene flake, connected by graphene leads. We assume that the
chemical potential in the external reservoirs and the leads is far
from the Dirac energy, while a gate fixes the chemical potential
of the quantum dot at the Dirac energy, $\epsilon = 0$. As there
are no extended states at $\epsilon = 0$ in the quantum dot,
transport processes take place through evanescent waves induced by
the contacts. This combination of chemical potentials is similar
to that considered in\cite{K06,TTTRB06} for the case of bulk
graphene.

We assume that the leads have a width $l$.  We also assume in the
following that the carrier concentration in the dots is
high\cite{K06,TTTRB06}, so that ${k_F}_L l \gg 1$, where ${k_F}_L$
is the Fermi wavelength in the leads. The transmission through the
constriction induces changes in the momentum parallel to the
interface, $k_y$, of order $\Delta k_y \sim l^{-1}$. As discussed
below, the transmission through the dot is determined by evanescent
waves with decay length $\kappa^{-1} \gtrsim l$. The allowed
parallel momentum of these modes is $k_y \sim \kappa \sim l^{-1}$.
Hence, the relevant incoming modes in the leads also must have $k_y
= {k_F}_L \sin ( \theta ) \approx l^{-1}$. The relevant states in
the leads are focused in the forward direction, $\theta \lesssim 1 /
( {k_F}_L l )$. Within these approximations, the incoming and
outgoing waves can be written as:
\begin{equation}
\Psi_{i} ( x_i ) \equiv \left( \begin{array}{c} {\Psi_A}_{i} ( x_i ) \\
{\Psi_B}_{i} ( x_i ) \end{array} \right) = \left( \begin{array}{c} 1 \\
\pm 1
\end{array} \right) e^{\pm i {k_F}_L x_{i}}
\label{incoming}
\end{equation}
where $x$ is the coordinate along the axis of the lead, the index $i
= in , out$ defines the incoming and outgoing leads, $A$ and $B$
label the two sublattices in graphene, and the two signs correspond
to waves moving in opposite directions. The states in
eq.(\ref{incoming}) are the same as those used in\cite{K06,TTTRB06}.
In the following, we will neglect inter-valley scattering in the
dot, so that the transmission through the dot can be analyzed for
each valley separately.

 The wave function in the incoming and outgoing leads, near
the contacts to the dot, $x_{i}=0$, can be written as:
\begin{eqnarray}
\Psi_{in} ( 0 ) &\equiv &\left( \begin{array}{c} 1 + R \\ 1 - R
\end{array} \right) \nonumber \\
\Psi_{out} ( 0 ) &\equiv &\left( \begin{array}{c} T \\ T
\end{array} \right)
\end{eqnarray}
where $R$ and $T$ are the reflection and transmission amplitudes,
and the carriers in the leads are supposed to be electrons
($\epsilon > 0$).

\begin{figure}
\begin{center}
\includegraphics*[width=8cm]{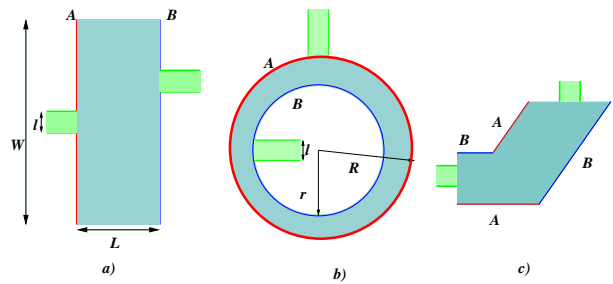}
\end{center}
\caption{(Color online). Sketch of the geometries of a graphene
quantum dot analyzed in the
  paper. a) Rectangular dot. b) Ring shaped dot. c) $60^\circ$ edge.}\label{conformal_dots}
\end{figure}

We assume that surface states exist at the edges of the dot. This is
the general case, as these states always exist when the two
sublattices of the honeycomb structure are not compensated at the
edge\cite{AB08}. The existence of these states leads to a depletion
of the density of extended states for energies $| \epsilon | \ll \vf
/ L$, where $L$ is a typical dimension of the dot. Hence, within
this range of energies we need only consider the midgap states
within the dot. Defining the complex variable $z = x + i y$ where
$x$ and $y$ are Cartesian coordinates, the wave function of the
states with zero energy must be of the form\cite{K06}:
\begin{eqnarray}
\Psi_K ( x , y ) &\equiv &\left( \begin{array}{c} \Psi^A_K ( x , y )
\\ \Psi^B_K ( x, y ) \end{array} \right)
\equiv \left( \begin{array}{c} f_K ( z ) \\
g_K (
\bar{z} ) \end{array} \right) \nonumber \\
\Psi_{K'} ( x , y ) &\equiv &\left( \begin{array}{c} \Psi^A_{K'} ( x
, y )
\\ \Psi^B_{K'} ( x, y ) \end{array} \right)
\equiv \left( \begin{array}{c} f_{K'} ( \bar{z} ) \\
g_{K'} ( z ) \end{array} \right)
\end{eqnarray}
where $f_K , g_K , f_{K'}$ and $g_{K'}$ are analytical functions,
and $K$ and $K'$ label the two inequivalent valleys in graphene. In
the following, we will consider only one valley. The results can be
extended to the other valley in a straightforward way. For the study
of the two cases shown in Fig.[\ref{conformal_dots}], we neglect
processes which lead to the hybridization of the two valleys. Note
that we are considering ballistic systems, where intervalley
scattering can be neglected. A different situation is that of a
$60^\circ$ edge, where intervalley scattering is crucial, even in
the ballistic limit.

\section{Results}
\subsection{Rectangular dot}
We first consider a rectangular dot, determined by its length $L$,
and width $W$, as sketched in Fig.[\ref{conformal_dots}a]. Each of
the leads be connected to an edge terminated in the $A$ sublattice,
and the other be connected to an edge terminated in the $B$
sublattice.

For simplicity, we assume periodic boundary conditions along the
vertical direction, as in Ref.\cite{K06}. For a graphene ribbon of
width $W \gg a$, the solutions for open boundary conditions can be
written as superpositions of waves with opposite momenta which are
possible solutions of an equivalent problem with periodic boundary
conditions. The difference between the two choices for a ribbon with
$N$ unit cells of length $a$ lies in the set of allowed momenta,
$k_\perp = n \pi / [( N + 1 ) a], n = 1 , \cdots , N$ for periodic
boundary conditions, and $k_\perp = \pm 2 \pi n / ( N a ) , n = 0 ,
\cdots , N/2$. If the transmissions to be calculated are a smooth
function of the momentum $k_\perp$, the results obtained using
either choice of boundary conditions converge to the same value for
$N \gg 1$.

For the rectangular dot considered here, the $x$ coordinate lies
within the range $- L / 2 \le x \le L / 2$, and the $y$ coordinate
can be written as $y = W \times \theta / ( 2 \pi )$, with $0 \le
\theta \le 2 \pi$. The wave functions inside the dot can be expanded
using the basis:
\begin{equation}
\Psi  ( x , y ) \equiv \sum_{n = - \infty}^{\infty} a_n \left(
\begin{array}{c}
e^{2 \pi n z / W} \\
0 \end{array} \right) + b_n \left( \begin{array}{c} 0 \\ e^{2 \pi n
\bar{z} / W} \end{array} \right) \label{wf}
\end{equation}
The incoming lead is attached at the position $z_1 = - L / 2$, and
the outgoing lead is attached to the position $z_2 = L / 2 + i W
\times \theta_0 / ( 2 \pi )$. The contact averages out the details
of the wave functions in the dot over a length of order $l$. Hence,
the description of the effects of the contact does not require the
infinite sum in Eq.(\ref{wf}). In the following, we include an upper
cutoff, $n_{max} \approx W / l$. The boundary conditions at $z_1$
and $z_2$ are:
\begin{eqnarray}
1 + R &= &\sum_{n = - n_{max}}^{n_{max}} a_n e^{2 \pi n z_1 / W} =
\sum_{n = - n_{max}}^{n_{max}} a_n e^{-\pi n L / W}
\nonumber \\
1 - R &= &\sum_{n = - n_{max}}^{n_{max}} b_n e^{2 \pi n \bar{z}_1 /
W}
= \sum_{n = - n_{max}}^{n_{max}} b_n e^{-\pi n L / W} \nonumber \\
T &= &\sum_{n = - n_{nmax}}^{n_{nmax}} a_n e^{2 \pi n z_2 / W} =
\sum_{n = - n_{max}}^{n_{max}} a_n e^{\pi n L / W} e^{i n \theta_0}
\nonumber \\
T &= &\sum_{n = - n_{max}}^{n_{max}} b_n e^{ 2 \pi n \bar{z}_2 / W}
= \sum_{n = - n_{max}}^{n_{max}} b_n e^{ \pi n L / W} e^{- i n
\theta_0} \nonumber \\
\end{eqnarray}
The zigzag boundary conditions at points at the edge with $x = -
L/2$ other than the position of the lead, $\theta=0$, and at the
edge with $x = L/2$ for $\theta \ne  \theta_0 $ read:
\begin{eqnarray}
\sum_{n = - n_{max}}^{n_{max}} b_n e^{- \pi n L/ W} e^{ - i n
\theta}
&= 0 &\, \, \, \, \, \, \, \theta \ne 0 \nonumber \\
\sum_{n = - n_{max}}^{n_{max}} a_n e^{\pi n L/ W} e^{ i n \theta} &=
0 &\, \, \, \, \, \, \, \theta \ne \theta_0
\end{eqnarray}
An Ansatz which is compatible with the two sets of boundary
conditions is
\begin{eqnarray}
a_n &= &T e^{- \pi n L / W} e^{- i n \theta_0} \nonumber \\
b_n &= &( 1 - R ) e^{\pi n L / W}
\end{eqnarray}
which leads to the equations
\begin{eqnarray}
1 + R &= &T \sum_{n = - n_{max}}^{n_{max}} e^{- 2 \pi n L / W} e^{- i n
  \theta_0} \nonumber \\
1 - R   &= &T \left[ \sum_{n = - n_{max}}^{n_{max}} e^{2 \pi n L /
W} \right]^{-1}
\end{eqnarray}
and
\begin{equation}
T = \frac{2}{\sum_{n = - n_{max}}^{n_{max}} e^{- 2 \pi n L / W} e^{-
i n \theta_0} + \left[ \sum_{n = - n_{max}}^{n_{max}} e^{2 \pi n L /
W} \right]^{-1}} \label{trans_coeff}
\end{equation}
The sums in Eq.(\ref{trans_coeff}) are dominated by the terms with
$n \sim \pm n_{max}$. Hence, we obtain:
\begin{equation}
| T |^2 \propto e^{- 4 \pi ( L / W ) ( W / l )} \label{trans_rec}
\end{equation}
Finally, when the two contacts are attached to the same type of
zigzag edge, the boundary conditions lead to either $a_n = 0$
(contacts attached to an $A$ terminated edge) or $b_n = 0$ (for a
$B$ terminated edge), and, as a result, $R=1$ and $T=0$.

The assumption of a rectangular shape implies that the top and
bottom boundaries may be of the armchair type, leading to
intervalley mixing\cite{BF06}. These boundaries are separated by a
length $W \gg l$. The valley mixing induces a change in the allowed
values of the transverse momentum, $k_y$, with respect to the case
of zigzag boundary conditions. The momentum shift is\cite{BF06}
$\Delta k_y \sim W^{-1}$. An incoming valley polarized mode can be
written as a superposition of exact eigenmodes in the presence of
valley mixing at the boundaries, with a spread in momentum of order
$\Delta k_y \sim l^{-1}$.

As discussed above, the transmission through the dot is determined
by evanescent waves with decay length $\kappa^{-1} \sim l$. The
parallel momentum associated to these states is $k_y \sim \kappa
\sim l^{-1}$. This superposition of valley polarized states will be
changed by the existence of valley mixing at the top and bottom
boundaries. The effects are of order $e^{- \Delta \kappa L} \sim
e^{- L / W}$, which are exponentially smaller than the transmission
coefficient obtained above, $| T | \sim e^{- 4 \pi L / l}$.

The previous analysis can be extended to a rectangular graphene
quantum dot in a constant magnetic field. The wave functions in
Eq.(\ref{wf}) in this case become
\begin{eqnarray}
\Psi ( x , y ) &\equiv &\sum_{n = -n_{max}}^{n_{max}} a_n \left(
\begin{array}{c}
    e^{2 \pi i n / W} e^{- x^2 / ( 2 l_B^2 )} \\ 0 \end{array} \right) + \nonumber \\ &+ &b_n
    \left( \begin{array}{c} 0 \\  e^{2 \pi i n / W} e^{ x^2 / ( 2 l_B^2 )}
    \end{array} \right)
\end{eqnarray}
where $l_B = \sqrt{\hbar c \ |e| B}$ is the magnetic length, $B$ is
the magnetic induction. The same manipulations described above allow
us to calculate the transmission coefficient, which turns out to be
unchanged with respect to Eq.(\ref{trans_rec}). The lack of
dependence of the transmission on the applied field is in agreement
with the insensitivity of the bulk transport to the magnetic field
when the chemical potential is at the Dirac point\cite{PSWG07}.

\subsection{Circular dot}
We now consider a ring-shaped dot, with the contacts attached to the
inner and outer edges, which are assumed to be $A$ and $B$
terminated, respectively\cite{WSG08}. As discussed there, a graphene
dot with an approximate circular shape can be bounded by different
edges where the majority sublattice is of either type. The advantage
of assuming the same boundary conditions throughout the edge is that
the problem retains circular symmetry, and the solutions have a well
defined angular momentum, $l$, where $l = 1 , \cdots l_{max} \approx
R / a$, where $R$ is the radius of the dot and $a$ is the lattice
constant. For $R / a  \gg 1$, superpositions of wavefunctions with
$l \gg 1$ can be built, with a small spread in angles, $\Psi ( r ,
\theta ) \ne 0$ only if $\theta_0 - \Delta \theta \le \theta \le
\theta_0 + \Delta \theta$, with $\Delta \theta \ll 2 \pi$. These
states are solutions of the Dirac equation at zero energy which are
not affected by the global properties of the boundaries of the dot.
Hence, they can be used to describe transport between positions at
the dot edges which are close enough so as to be insensitive to the
global properties of the edges. The case when the transport
properties are influenced by the type of edges requires input about
features at distances comparable to the lattice spacing, and will be
considered in the next subsection.

The outer and inner radii are $R_1$ and $R_2$, as shown in
Fig.[\ref{conformal_dots}b]. The modes at $\epsilon = 0$ inside the
dot which satisfy the boundary conditions can be obtained by a
conformal transformation which changes the rectangular (cylindrical)
dot considered in the previous subsection into a ring. This
transformation is
\begin{equation}
w ( z ) = \sqrt{R_1 R_2}  e^{2 \pi i z / W}
\end{equation}
and $e^{2 \pi L / W} = R_1 / R_2$. Using this transformation, the
wave functions in Eq.(\ref{wf}) become
\begin{equation}
\Psi  ( x , y ) \equiv \sum_{n = - \infty}^{\infty} a_n \left(
\begin{array}{c}
z^n \\
0 \end{array} \right) + b_n \left( \begin{array}{c} 0 \\ \bar{z}^n
\end{array} \right) \label{wf_2}
\end{equation}
The size of the contact, $l$, induces a maximum value $n$,
$n_{max} \sim 2 \pi R_2 / l$. From the wave functions in
Eq.(\ref{wf_2}), and using the same analysis as for the
rectangular dot in the previous subsection, we wind, in analogy
with Eq.(\ref{trans_rec}):
\begin{equation}
| T |^2 \sim \left( \frac{R_2}{R_1} \right)^{8 \pi R_2/l}
\label{trans_circle}
\end{equation}
If we write $2 \pi R_2 = W$ and $R_1 \approx R_2 + L$, this
expression reduces to Eq.(\ref{trans_rec}) when $R_1 \sim R_2 \gg
R_1 - R_2$.

As in the case of the rectangular dot, the only allowed solution
when the two contacts are attached to the same boundary is $R=1$
and $T=0$.

Dots of different shapes can be obtained using other conformal
transformations. For instance, the circular dot analyzed here can
be turned into a dot with a corrugated boundary using the mapping
\begin{equation}
w' ( w ) = w + \lambda w^m
\end{equation}
where $m$ fixes the period of the corrugation, and $\lambda \propto \delta
R / R^m$ gives its
amplitude. The transmission coefficient in eq.(\ref{trans_circle}) acquires
corrections of order $( R / l ) ( \delta R / R )^m$.

This analysis can also be extended to a finite magnetic field. In
that case the wave functions are
\begin{eqnarray}
\Psi ( x , y ) &\equiv &\sum_{n = -n_{max}}^{n_{max}} a_n \left(
\begin{array}{c}
    z^n e^{- r^2 / ( 2 l_B^2 )} \\ 0 \end{array} \right) + \nonumber \\ &+ &b_n
    \left( \begin{array}{c} 0 \\ \bar{z}^n e^{ r^2 / ( 2 l_B^2 )}
    \end{array} \right)
\end{eqnarray}
The boundary conditions imply
\begin{eqnarray}
1+R &= &e^{- R_1^2 / ( 2 l_B^2 )} \sum_{n = -n_{max}}^{n_{max}} a_n R_1^n
  \nonumber \\
1-R &= &e^{ R_1^2 / ( 2 l_B^2 )} \sum_{n = -n_{max}}^{n_{max}} b_n
R_1^n
\nonumber \\
T &= &e^{- R_2^2 / ( 2 l_B^2 )} \sum_{n = -n_{max}}^{n_{max}} a_n R_2^n
e^{i n \theta_0}  \nonumber \\
T &= &e^{ R_2^2 / ( 2 l_B^2 )} \sum_{n = -n_{max}}^{n_{max}} b_n
R_2^n e^{-i
  n \theta_0}
\end{eqnarray}
A possible solution is of the form
\begin{eqnarray}
b_n &= &\frac{C}{R_2^n} e^{i n \theta_0} \nonumber \\
a_n &= &\frac{A}{R_1^n}
\end{eqnarray}
leading to the equations
\begin{eqnarray}
1+R &= &A e^{- R_1^2 / ( 2 l_B^2 )} n_{max} \nonumber \\
1-R &= &C e^{ R_1^2 / ( 2 l_B^2 )} \left( \frac{R_1}{R_2}
\right)^{n_{max}}
\nonumber \\
T &= &A e^{- R_2^2 / ( 2 l_B^2 )} \left( \frac{R_2}{R_1}
\right)^{n_{max}}
e^{-i n_{max} \theta_0} \nonumber \\
T &= &C e^{ R_2^2 / ( 2 l_B^2 )} n_{max}
\end{eqnarray}
Thus, the transmission coefficient is
\begin{eqnarray}
\left| T \right|^2 &= &n_{max}^2 \left( \frac{R_2}{R_1} \right)^{2
n_{max}} e^{- (R_1^2 -
  R_2^2 ) / l_B^2 } = \nonumber \\ &= &n_{max}^2 \left( \frac{R_2}{R_1} \right)^{2 n_{max}}
  e^{- \Phi / \Phi_0}
\end{eqnarray}
where $\Phi$ is the magnetic flux per ring and $ \Phi_0 = \pi \hbar
c / |e|$ is the flux quantum. The dependence of transport through
the quantum dot on the magnetic field is exponential, rather than
oscillatory as in case of the standard Aharonov-Bohm
effect\cite{BH91}, which reflects a specifics of transport via {\it
evanescent} waves.

Thus, the conductance of circular quantum dot turns out to be very
sensitive to the flux through the ring. This can be potentially
interesting in light of development of magnetic sensors for
measurements of very low fields without use of superconductivity.
\begin{figure}
\begin{center}
\includegraphics*[width=8cm]{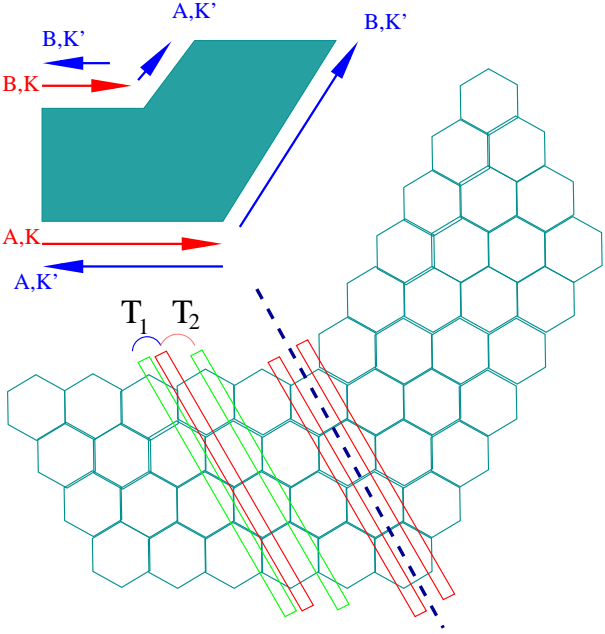}
\end{center}
\caption{(Color online). Sketch of the $60^\circ$ junction analyzed
in the text. Top left: an
  incoming wave from the left, built up from eigenstates in the vicinity of
  the $K$ valley has to be either transmitted or reflected as a superposition
  of states from the $K'$ valley. Bottom: Scheme used to analyze the
  transmission in the honeycomb lattice. The junction has a reflection
  symmetry around its center, shown as dashed line. On each side, the nearest
  neighbor hopping problem can be written as a sum of coupling between
  nearest neighbor transverse stripes. See text for details.}\label{edge_60}
\end{figure}
\subsection{Corner between two facets}
We finally consider a 60$^\circ$ angle boundary between an edge with
$A$ termination, and an edge with $B$ termination, as sketched in
Fig.[\ref{conformal_dots}c]. The leads are attached at distances
$l_1$ and $l_2$ from the corner. The transmission along a ribbon
with this geometry has been analyzed numerically in
Refs.\cite{RB07,R07,ILFB08}.  A full solution cannot be obtained
within the continuum approximation, as in the previous cases, since
both transmission and reflection at the wedge require Umklapp
processes changing the valley index of the incoming electron. In
this respect, the problem is similar to that of the transmission
across a p-n junction near the Dirac point\cite{ABRB07}.

We consider the ribbon with an angle shown in Fig.[\ref{edge_60}].
The system has reflection symmetry around an axis which connects
the vortices of the angles at both sides. Far away from the
corner, the system reduces to a zigzag ribbon. We assume that the
edges of the ribbon are at $y = \pm W/2$, and $W \gg a$, where $a$
is the lattice constant. The wave function of an incoming wave can
be written as
\begin{equation}
\Psi_K^{in} ( x , y ) \equiv \left( \begin{array}{c}  \sinh \left[ \kappa \left( y - \frac{W}{2} \right) \right] \\
      - \sqrt{\frac{k + \kappa}{k -
      \kappa}} e^{\kappa ( y-W/2 )} + \sqrt{\frac{k-\kappa}{k+\kappa}}
  e^{\kappa ( W/2-y)} \end{array} \right)
e^{i k x} \label{incoming_ev}\end{equation} where $\kappa$ and the
energy of the state are given by:
\begin{eqnarray}
\kappa &= &k \tanh \left( \kappa W \right) \nonumber \\
\epsilon_k &= &\vf \sqrt{k^2 - \kappa^2}
\end{eqnarray}
For $k W \gg 1$, we find
\begin{eqnarray}
\kappa &\approx &k ( 1 - e^{- 2 k W} + \cdots ) \nonumber \\
\epsilon_k &\approx &\vf k e^{- k W}
\end{eqnarray}
The solution (\ref{incoming_ev}) is valid for $k \ge W^{-1}$.
\begin{figure}
\begin{center}
\includegraphics*[width=5cm]{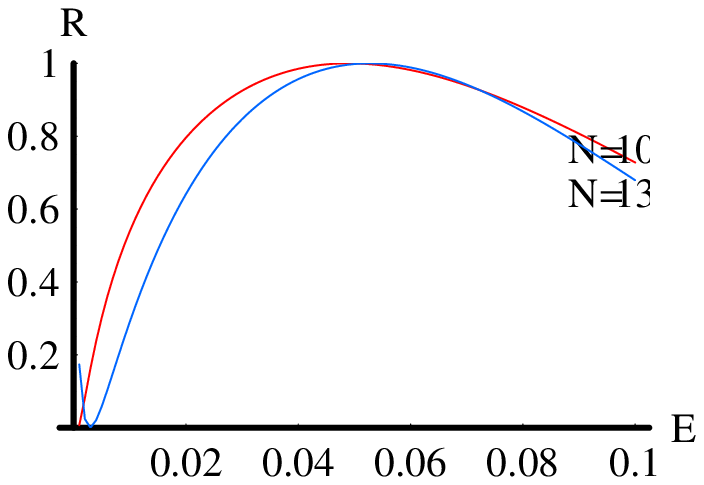}
\includegraphics*[width=5cm]{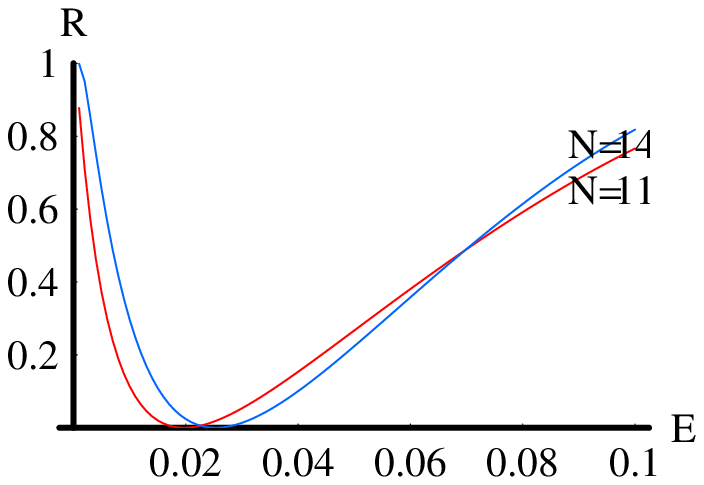}
\includegraphics*[width=5cm]{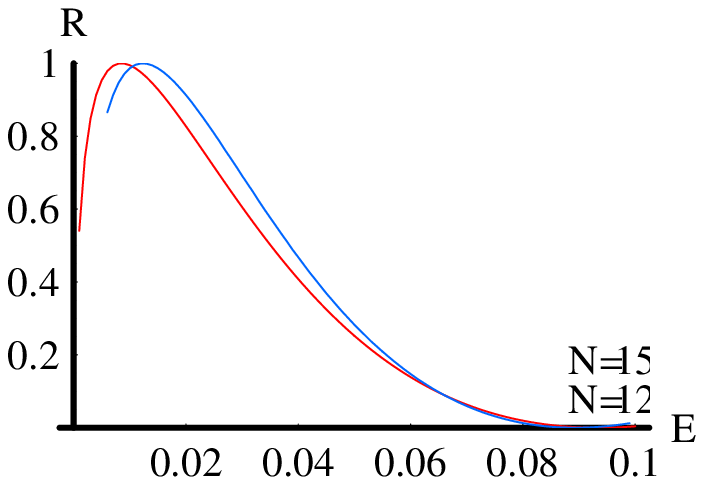}
\end{center}
\caption{(Color online). Reflection at a 60$^\circ$ junction like
the one sketched in Fig.[\protect{\ref{edge_60}}] for different
junction widths. $N$ gives the number of unit cells across the
junction. The width of each ribbon which make up the junction is $3
/ 2 a N$, where $a$ is the length of the C-C bond. The energy is in
units of the nearest neighbor hopping.}\label{edge_60_res}
\end{figure}
\begin{figure}
\begin{center}
\includegraphics*[width=8cm]{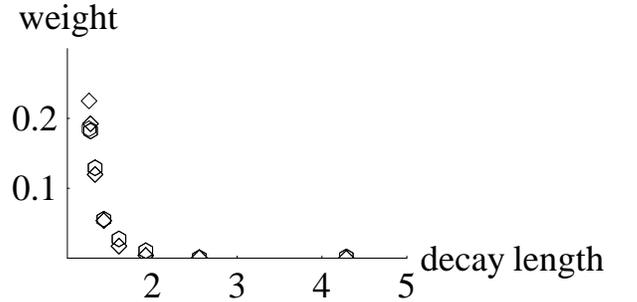}
\end{center}
\caption{(Color online). Weights of the wavefunction passing through
the junction sketched in Fig.[\protect{\ref{edge_60}}] on evanescent
waves of different decay lengths. The width of the junction is 16
unit cells, and there are 30 evanescent waves. The decay length is
in units of the lattice spacing in the direction parallel to the
edges ($\sqrt{3} \times a$, where $a$ is the C-C bond). Diamonds
show waves symmetric around the junction, and hexagons show
antisymmetric waves (note that the waves studied here decay as
function of the distance to the junction, not to the edges of the
ribbon).}\label{weights}
\end{figure}

The reflected and transmitted waves can be written as in
Eq.(\ref{incoming_ev}), except that, for the reflected wave the
momentum is reversed, and the wave packet is built up from electrons
at the $K'$ valley:
\begin{equation}
\Psi_{K'}^{ref} ( x , y ) \equiv \left( \begin{array}{c}  \sinh \left[ \kappa
      \left( y - \frac{W}{2} \right) \right] \\
      - \sqrt{\frac{k + \kappa}{k -
      \kappa}} e^{\kappa ( y-W/2 )} + \sqrt{\frac{k-\kappa}{k+\kappa}}
  e^{\kappa ( W/2-y)} \end{array} \right)
e^{- i k x} \label{incoming_ev2}
\end{equation}
This wave function has to be matched to that in
Eq.(\ref{incoming_ev}) at the interface. In the continuum limit,
the overlap between wave functions at $K$ and $K'$ valleys is
zero, and the matching cannot be carried out. We analyze the
corrections induced by the finite lattice by dividing each ribbon
reaching the junction into stripes, as sketched in
Fig.[\ref{edge_60}]. Using a nearest-neighbor tight-binding model,
each one-dimensional ribbon is coupled to its two nearest
neighbors, and the Hamiltonian can be written as a sum of $N
\times N$ terms, $T_1$ and $T_2$, where $N$ is the number of atoms
within each ribbon:
\begin{eqnarray}
\epsilon_k \alpha_i^M &= &\sum_{j=1}^N T^1_{ij} \beta_j^M +
T^2_{ij}
\beta_j^{M+1} \, \, \, \, \, \, \, \, \, \, i=1, \cdots , N \nonumber \\
\epsilon_k\beta_i^M &= &\sum_{j=1}^N T^1_{ij} \alpha_j^M +
T^2_{ij} \alpha_j^{M-1} \, \, \, \, \, \, \, \, \, \, i=1, \cdots
, N \label{amplitudes}
\end{eqnarray}
where $M$ is a stripe index.

Using the inversion symmetry of the junction, we can analyze
separately states which are even and odd with respect to the axis
which joins the two angles of the junction. Within each subsector,
the problem is reduced to the reflection of an incoming wave of
energy $\epsilon_k$ at a sharp boundary at the location of the
junction. The coupling between the stripes at each side of the
junction becomes, in the new geometry, an energy shift equal to
the hopping $t$ at the atoms in the last stripe of the system:
\begin{equation}
\epsilon_k \alpha_i^0 = \pm t \alpha_i^0 +  \sum_{j=1}^N T^1_{ij} \beta_j^0 \,
\, \, \, \, \, \, \, \, \, i=1, \cdots , N
\label{boundary}
\end{equation}
An incoming wave of momentum $k$ and energy $\epsilon_k$ defines a
set of coefficients, $\{ \alpha_i^k \} , \{ \beta_i^k \}$, such
that the Ansatz:
\begin{eqnarray}
\alpha_i^M &= &\alpha_i^k e^{i k M} \nonumber \\
\beta_i^M &= &\beta_i^k e^{i k M}
\label{propagating}
\end{eqnarray}
is a solution of Eq.(\ref{amplitudes}). In the limit $N \gg 1$,
and $\epsilon_k \ll t$, these amplitudes are well approximated by
Eq.(\ref{incoming_ev}). For a given energy $\epsilon_k\sim t
e^{-N}$, we can define only one incoming and outgoing set of
amplitudes  as in Eq.(\ref{propagating}).  We can also define
$N-1$ amplitudes $\{ \alpha_i^\kappa \}, \{ \beta_i^\kappa \}$
such that
\begin{eqnarray}
\alpha_i^M &= &\alpha_i^k e^{- \kappa M} \nonumber \\
\beta_i^M &= &\beta_i^k e^{- \kappa M}
\label{evanescent}
\end{eqnarray}
satisfy Eq.(\ref{amplitudes}). For energies very near the Dirac
point, the solutions (\ref{propagating}) and (\ref{evanescent})
are such that either the amplitudes $ \alpha_i^k  ,
\alpha_i^\kappa $ or  $ \beta_j^k  ,  \beta_j^\kappa $ are
exponentially small when $T^1_{i,j} , T^2_{ij} \ne 0$.

The boundary condition at the edge of the system is given by
Eq.(\ref{boundary}) where the amplitudes can be expanded into an
incoming, a reflected, and $M_1$ evanescent waves:
\begin{eqnarray}
\alpha_i^0 &= \alpha_i^k + R \alpha_i^{-k} + \sum_{\kappa} C_\kappa \alpha_i^\kappa
\nonumber \\
\beta_i^0 &= \beta_i^k + R \beta_i^{-k} + \sum_{\kappa} C_\kappa
\beta_i^\kappa
\label{variables}
\end{eqnarray}
The insertion of these expressions into Eq.(\ref{boundary}) leads
to $N$ equations with the $N-1$ unknowns $\{ C_\kappa \}$, and the
reflection coefficient $R$. The even and odd combinations of the
initial junction problem allows us to define to reflection
coefficients, $R_{\pm}$ obtained using the two possible signs in
Eq.(\ref{boundary}). It is easy to show that the transmission
coefficient of the junction can be written as $T = ( R_+ - R_- ) /
2$.

The only solution of Eq.(\ref{boundary}) at low energies requires
all the amplitudes $\alpha_i^0$ to be exponentially small, as a
finite set of values $\alpha_i^0$ and $\beta_j^0$ are incompatible
if $T^1_{ij} \ne0$. Hence, the parameters $C_\kappa^\pm$ and
$R_\pm$ are determined by a set of equations which are independent
of the choice of sign in Eq.(\ref{boundary}), so that $R_+ \approx
R_-$ with exponential accuracy. Hence, the transmission
coefficient of the original junction also vanishes with
exponential accuracy, in agreement with the numerical calculations
in Refs.\cite{RB07,R07}.

Results are shown in Fig.[\ref{edge_60_res}]. They are consistent
with those reported in\cite{ILFB08}. The dependence on energy of the
reflection coefficient shows three characteristic patterns, which
are repeated as function of the width of the junction, as shown in
the Figure. This approximate periodicity is reminiscent of the
alternance of metallic and non metallic features in carbon nanotubes
and graphene nanoribbons.

 The calculation
described here requires the existence of evanescent waves which have
a finite overlap with wavepackets derived from both the $K$ and $K'$
points of the Brillouin zone, in a similar way to the transmission
problem analyzed in Ref.\cite{ABRB07}. The coupling between the two
valleys can be analyzed in detail by calculating the relative
weights of the different evanescent waves which must be defined at
the junction. Results for a junction of width $N=16$ are shown in
Fig.[\ref{weights}], where the decay length is defined as $l_{ev} =
1/\kappa$ in eq.(\ref{evanescent}). Most of the weight is
concentrated on evanescent waves with a short decay length, which
cannot be ascribed to a given valley.

\section{Conclusions}
We have studied the transmission, at zero energy, through graphene
quantum dots attached to leads with one incoming and outgoing
channels, extending the analysis in Refs.\cite{K06,TTTRB06}. We
assume that, for energies at distances to the Dirac point smaller
than $\vf / L$, where $L$ is the typical dimension of the system,
extended states can be neglected, and the electronic properties
are mainly determined by evanescent waves, or by localized states
induced by boundaries.

We have shown that the dots of various shapes can be analyzed,
using conformal transformations which preserve the nature of the
states at zero energy. An exception is an angle between zigzag
boundaries with different terminations, where the solution of the
problem requires the analysis of evanescent waves which do not
have a well defined valley index, and cannot be described by the
continuum Dirac equation.

We find that:

(i) The transmission vanishes when the leads are attached to the
same edge.

(ii) The transmission depends exponentially on the ratio between
the size of the dot and the width of the contact.

(iii) The shape of the dot changes significantly the transmission.

(iv) The effects of a magnetic field are strongly dependent on the
shape of the dot. A rectangular dot is unique in that a magnetic
field does not change the transport properties (see
also\cite{PSWG07}). In general, the conduction of the dot is very
sensitive to the magnetic flux through it. A similar behavior for
dots in the diffusive regime has been reported in\cite{AL08}.

(v) The transmission when the two contacts are attached to an edge
such that both the transmitted and reflected wave belong to the
opposite valley as the incoming wave depends on the lattice
structure at short distances.

\section{Acknowledgements}
This work was supported by MEC (Spain) through grant
FIS2005-05478-C02-01 and CONSOLIDER CSD2007-00010, the Comunidad de
Madrid, through the program CITECNOMIK, CM2006-S-0505-ESP-0337, the
European Union Contract 12881 (NEST), and the Stichting voor
Fundamenteel Onderzoek der Materie (FOM) (the Netherlands). We
appreciate helpful conversations with L. Brey and M. A. H.
Vozmediano.

\bibliography{dots_zero_energy}

\newcommand{\npb}{Nucl. Phys. B}\newcommand{\adv}{Adv.
  Phys.}\newcommand{\epl}{Europhys. Lett.}
\begin{thebibliography}{29}
\expandafter\ifx\csname natexlab\endcsname\relax\def\natexlab#1{#1}\fi
\expandafter\ifx\csname bibnamefont\endcsname\relax
  \def\bibnamefont#1{#1}\fi
\expandafter\ifx\csname bibfnamefont\endcsname\relax
  \def\bibfnamefont#1{#1}\fi
\expandafter\ifx\csname citenamefont\endcsname\relax
  \def\citenamefont#1{#1}\fi
\expandafter\ifx\csname url\endcsname\relax
  \def\url#1{\texttt{#1}}\fi
\expandafter\ifx\csname urlprefix\endcsname\relax\def\urlprefix{URL }\fi
\providecommand{\bibinfo}[2]{#2}
\providecommand{\eprint}[2][]{\url{#2}}

\bibitem[{\citenamefont{Novoselov et~al.}(2004)\citenamefont{Novoselov, Geim,
  Morozov, Jiang, Zhang, Dubonos, Grigorieva, and Firsov}}]{Netal04}
\bibinfo{author}{\bibfnamefont{K.~S.} \bibnamefont{Novoselov}},
  \bibinfo{author}{\bibfnamefont{A.~K.} \bibnamefont{Geim}},
  \bibinfo{author}{\bibfnamefont{S.~V.} \bibnamefont{Morozov}},
  \bibinfo{author}{\bibfnamefont{D.}~\bibnamefont{Jiang}},
  \bibinfo{author}{\bibfnamefont{Y.}~\bibnamefont{Zhang}},
  \bibinfo{author}{\bibfnamefont{S.~V.} \bibnamefont{Dubonos}},
  \bibinfo{author}{\bibfnamefont{I.~V.} \bibnamefont{Grigorieva}},
  \bibnamefont{and} \bibinfo{author}{\bibfnamefont{A.~A.}
  \bibnamefont{Firsov}}, \bibinfo{journal}{Science}
  \textbf{\bibinfo{volume}{306}}, \bibinfo{pages}{666} (\bibinfo{year}{2004}).

\bibitem[{\citenamefont{Novoselov et~al.}(2005)\citenamefont{Novoselov, Jiang,
  Schedin, Booth, Khotkevich, Morozov, and Geim}}]{Netal05b}
\bibinfo{author}{\bibfnamefont{K.~S.} \bibnamefont{Novoselov}},
  \bibinfo{author}{\bibfnamefont{D.}~\bibnamefont{Jiang}},
  \bibinfo{author}{\bibfnamefont{F.}~\bibnamefont{Schedin}},
  \bibinfo{author}{\bibfnamefont{T.~J.} \bibnamefont{Booth}},
  \bibinfo{author}{\bibfnamefont{V.~V.} \bibnamefont{Khotkevich}},
  \bibinfo{author}{\bibfnamefont{S.~V.} \bibnamefont{Morozov}},
  \bibnamefont{and} \bibinfo{author}{\bibfnamefont{A.~K.} \bibnamefont{Geim}},
  \bibinfo{journal}{Proc. Nat. Acad. Sc.} \textbf{\bibinfo{volume}{102}},
  \bibinfo{pages}{10451} (\bibinfo{year}{2005}).

\bibitem[{\citenamefont{Geim and Novoselov}(2007)}]{GN07}
\bibinfo{author}{\bibfnamefont{A.~K.} \bibnamefont{Geim}} \bibnamefont{and}
  \bibinfo{author}{\bibfnamefont{K.~S.} \bibnamefont{Novoselov}},
  \bibinfo{journal}{Nature Materials} \textbf{\bibinfo{volume}{6}},
  \bibinfo{pages}{183} (\bibinfo{year}{2007}).

\bibitem[{\citenamefont{Katsnelson and Novoselov}(2007)}]{KN07}
\bibinfo{author}{\bibfnamefont{M.~I.} \bibnamefont{Katsnelson}}
  \bibnamefont{and} \bibinfo{author}{\bibfnamefont{K.~S.}
  \bibnamefont{Novoselov}}, \bibinfo{journal}{Sol. State Commun.}
  \textbf{\bibinfo{volume}{143}}, \bibinfo{pages}{3} (\bibinfo{year}{2007}).

\bibitem[{\citenamefont{{Castro Neto} et~al.}(2007)\citenamefont{{Castro Neto},
  Guinea, Peres, Novoselov, and Geim}}]{NGPNG07}
\bibinfo{author}{\bibfnamefont{A.~H.} \bibnamefont{{Castro Neto}}},
  \bibinfo{author}{\bibfnamefont{F.}~\bibnamefont{Guinea}},
  \bibinfo{author}{\bibfnamefont{N.~M.~R.} \bibnamefont{Peres}},
  \bibinfo{author}{\bibfnamefont{K.~S.} \bibnamefont{Novoselov}},
  \bibnamefont{and} \bibinfo{author}{\bibfnamefont{A.~K.} \bibnamefont{Geim}}
  (\bibinfo{year}{2007}), \bibinfo{note}{{Rev. Mod. Phys., in press}},
  \eprint{arXiv:0709.1163}.

\bibitem[{\citenamefont{Katsnelson}(2006)}]{K06}
\bibinfo{author}{\bibfnamefont{M.~I.} \bibnamefont{Katsnelson}},
  \bibinfo{journal}{Eur. J. Phys. B} \textbf{\bibinfo{volume}{51}},
  \bibinfo{pages}{157} (\bibinfo{year}{2006}).

\bibitem[{\citenamefont{Tworzydo et~al.}(2006)\citenamefont{Tworzydo,
  Trauzettel, Titov, Rycerz, and Beenakker}}]{TTTRB06}
\bibinfo{author}{\bibfnamefont{J.}~\bibnamefont{Tworzydo}},
  \bibinfo{author}{\bibfnamefont{B.}~\bibnamefont{Trauzettel}},
  \bibinfo{author}{\bibfnamefont{M.}~\bibnamefont{Titov}},
  \bibinfo{author}{\bibfnamefont{A.}~\bibnamefont{Rycerz}}, \bibnamefont{and}
  \bibinfo{author}{\bibfnamefont{C.~W.~J.} \bibnamefont{Beenakker}},
  \bibinfo{journal}{Phys. Rev. Lett.} \textbf{\bibinfo{volume}{96}},
  \bibinfo{pages}{246802} (\bibinfo{year}{2006}).

\bibitem[{\citenamefont{Miao et~al.}(2007)\citenamefont{Miao, Wijeratne, Zhang,
  Coskun, Bao, and Lau}}]{Metal07}
\bibinfo{author}{\bibfnamefont{F.}~\bibnamefont{Miao}},
  \bibinfo{author}{\bibfnamefont{S.}~\bibnamefont{Wijeratne}},
  \bibinfo{author}{\bibfnamefont{Y.}~\bibnamefont{Zhang}},
  \bibinfo{author}{\bibfnamefont{U.~C.} \bibnamefont{Coskun}},
  \bibinfo{author}{\bibfnamefont{W.}~\bibnamefont{Bao}}, \bibnamefont{and}
  \bibinfo{author}{\bibfnamefont{C.~N.} \bibnamefont{Lau}},
  \bibinfo{journal}{Science} \textbf{\bibinfo{volume}{317}},
  \bibinfo{pages}{1530} (\bibinfo{year}{2007}).

\bibitem[{\citenamefont{Danneau et~al.}(2008)\citenamefont{Danneau, Wu,
  Craciun, Russo, Tomi, Salmilehto, Morpurgo, and Hakonen}}]{Detal08}
\bibinfo{author}{\bibfnamefont{R.}~\bibnamefont{Danneau}},
  \bibinfo{author}{\bibfnamefont{F.}~\bibnamefont{Wu}},
  \bibinfo{author}{\bibfnamefont{M.~F.} \bibnamefont{Craciun}},
  \bibinfo{author}{\bibfnamefont{S.}~\bibnamefont{Russo}},
  \bibinfo{author}{\bibfnamefont{M.~Y.} \bibnamefont{Tomi}},
  \bibinfo{author}{\bibfnamefont{J.}~\bibnamefont{Salmilehto}},
  \bibinfo{author}{\bibfnamefont{A.~F.} \bibnamefont{Morpurgo}},
  \bibnamefont{and} \bibinfo{author}{\bibfnamefont{P.~J.}
  \bibnamefont{Hakonen}} (\bibinfo{year}{2008}), \eprint{arXiv:0711.4306}.

\bibitem[{\citenamefont{Fujita et~al.}(1996)\citenamefont{Fujita, Wakabayashi,
  Nakada, and Kusakabe}}]{FWNK96}
\bibinfo{author}{\bibfnamefont{M.}~\bibnamefont{Fujita}},
  \bibinfo{author}{\bibfnamefont{K.}~\bibnamefont{Wakabayashi}},
  \bibinfo{author}{\bibfnamefont{K.}~\bibnamefont{Nakada}}, \bibnamefont{and}
  \bibinfo{author}{\bibfnamefont{K.}~\bibnamefont{Kusakabe}},
  \bibinfo{journal}{J. Phys. Soc. Jpn.} \textbf{\bibinfo{volume}{65}},
  \bibinfo{pages}{1920} (\bibinfo{year}{1996}).

\bibitem[{\citenamefont{Vozmediano et~al.}(2005)\citenamefont{Vozmediano,
  L\'opez-Sancho, Stauber, and Guinea}}]{VLSG05}
\bibinfo{author}{\bibfnamefont{M.~A.} \bibnamefont{Vozmediano}},
  \bibinfo{author}{\bibfnamefont{M.~P.} \bibnamefont{L\'opez-Sancho}},
  \bibinfo{author}{\bibfnamefont{T.}~\bibnamefont{Stauber}}, \bibnamefont{and}
  \bibinfo{author}{\bibfnamefont{F.}~\bibnamefont{Guinea}},
  \bibinfo{journal}{Phys. Rev. B} \textbf{\bibinfo{volume}{72}},
  \bibinfo{pages}{155121} (\bibinfo{year}{2005}).

\bibitem[{\citenamefont{Pereira et~al.}(2006)\citenamefont{Pereira, Guinea,
  {Lopes dos Santos}, Peres, and {Castro Neto}}}]{PGLPN06}
\bibinfo{author}{\bibfnamefont{V.~M.} \bibnamefont{Pereira}},
  \bibinfo{author}{\bibfnamefont{F.}~\bibnamefont{Guinea}},
  \bibinfo{author}{\bibfnamefont{J.~M.} \bibnamefont{{Lopes dos Santos}}},
  \bibinfo{author}{\bibfnamefont{N.~M.} \bibnamefont{Peres}}, \bibnamefont{and}
  \bibinfo{author}{\bibfnamefont{A.~H.} \bibnamefont{{Castro Neto}}},
  \bibinfo{journal}{Phys. Rev. Lett.} \textbf{\bibinfo{volume}{96}},
  \bibinfo{pages}{036801} (\bibinfo{year}{2006}).

\bibitem[{\citenamefont{Akhmerov and Beenakker}(2008)}]{AB08}
\bibinfo{author}{\bibfnamefont{A.~R.} \bibnamefont{Akhmerov}} \bibnamefont{and}
  \bibinfo{author}{\bibfnamefont{C.~W.~J.} \bibnamefont{Beenakker}},
  \bibinfo{journal}{Phys. Rev. B} \textbf{\bibinfo{volume}{77}},
  \bibinfo{pages}{085423} (\bibinfo{year}{2008}).

\bibitem[{\citenamefont{Guinea et~al.}(2008)\citenamefont{Guinea, Katsnelson,
  and Vozmediano}}]{GKV07}
\bibinfo{author}{\bibfnamefont{F.}~\bibnamefont{Guinea}},
  \bibinfo{author}{\bibfnamefont{M.~I.} \bibnamefont{Katsnelson}},
  \bibnamefont{and} \bibinfo{author}{\bibfnamefont{M.~A.~H.}
  \bibnamefont{Vozmediano}}, \bibinfo{journal}{Phys. Rev. B}
  \textbf{\bibinfo{volume}{77}}, \bibinfo{pages}{075422}
  (\bibinfo{year}{2008}).

\bibitem[{\citenamefont{Louis et~al.}(2007)\citenamefont{Louis, Verg\'es,
  Guinea, and Chiappe}}]{LVGC07}
\bibinfo{author}{\bibfnamefont{E.}~\bibnamefont{Louis}},
  \bibinfo{author}{\bibfnamefont{J.~A.} \bibnamefont{Verg\'es}},
  \bibinfo{author}{\bibfnamefont{F.}~\bibnamefont{Guinea}}, \bibnamefont{and}
  \bibinfo{author}{\bibfnamefont{G.}~\bibnamefont{Chiappe}},
  \bibinfo{journal}{Phys. Rev. B} \textbf{\bibinfo{volume}{75}},
  \bibinfo{pages}{085440} (\bibinfo{year}{2007}).

\bibitem[{\citenamefont{Titov}(2007)}]{T07}
\bibinfo{author}{\bibfnamefont{M.}~\bibnamefont{Titov}},
  \bibinfo{journal}{Europhys. Lett.} \textbf{\bibinfo{volume}{79}},
  \bibinfo{pages}{17004} (\bibinfo{year}{2007}).

\bibitem[{\citenamefont{Wunsch et~al.}(2008)\citenamefont{Wunsch, Stauber, and
  Guinea}}]{WSG08}
\bibinfo{author}{\bibfnamefont{B.}~\bibnamefont{Wunsch}},
  \bibinfo{author}{\bibfnamefont{T.}~\bibnamefont{Stauber}}, \bibnamefont{and}
  \bibinfo{author}{\bibfnamefont{F.}~\bibnamefont{Guinea}},
  \bibinfo{journal}{Phys. Rev. B} \textbf{\bibinfo{volume}{77}},
  \bibinfo{pages}{035316} (\bibinfo{year}{2008}).

\bibitem[{\citenamefont{Bunch et~al.}(2005)\citenamefont{Bunch, Yaish, Brink,
  Bolotin, and McEuen}}]{Betal05}
\bibinfo{author}{\bibfnamefont{J.~S.} \bibnamefont{Bunch}},
  \bibinfo{author}{\bibfnamefont{Y.}~\bibnamefont{Yaish}},
  \bibinfo{author}{\bibfnamefont{M.}~\bibnamefont{Brink}},
  \bibinfo{author}{\bibfnamefont{K.}~\bibnamefont{Bolotin}}, \bibnamefont{and}
  \bibinfo{author}{\bibfnamefont{P.~L.} \bibnamefont{McEuen}},
  \bibinfo{journal}{Nano Lett.} \textbf{\bibinfo{volume}{5}},
  \bibinfo{pages}{2887} (\bibinfo{year}{2005}).

\bibitem[{\citenamefont{Han et~al.}(2007)\citenamefont{Han, \"Ozyilmaz, Zhang,
  and Kim}}]{HOZK07}
\bibinfo{author}{\bibfnamefont{M.~Y.} \bibnamefont{Han}},
  \bibinfo{author}{\bibfnamefont{B.}~\bibnamefont{\"Ozyilmaz}},
  \bibinfo{author}{\bibfnamefont{Y.}~\bibnamefont{Zhang}}, \bibnamefont{and}
  \bibinfo{author}{\bibfnamefont{P.}~\bibnamefont{Kim}},
  \bibinfo{journal}{Phys. Rev. Lett.} \textbf{\bibinfo{volume}{98}},
  \bibinfo{pages}{206805} (\bibinfo{year}{2007}).

\bibitem[{\citenamefont{Avouris et~al.}(2007)\citenamefont{Avouris, Chen, and
  Perebeinos}}]{AZP07}
\bibinfo{author}{\bibfnamefont{P.}~\bibnamefont{Avouris}},
  \bibinfo{author}{\bibfnamefont{Z.}~\bibnamefont{Chen}}, \bibnamefont{and}
  \bibinfo{author}{\bibfnamefont{V.}~\bibnamefont{Perebeinos}},
  \bibinfo{journal}{Nature Nanotechnology} \textbf{\bibinfo{volume}{2}},
  \bibinfo{pages}{605} (\bibinfo{year}{2007}).

\bibitem[{\citenamefont{Ponomarenko et~al.}(2008)\citenamefont{Ponomarenko,
  Schedin, Katsnelson, Yang, Hill, Novoselov, and K.Geim}}]{Petal08}
\bibinfo{author}{\bibfnamefont{L.~A.} \bibnamefont{Ponomarenko}},
  \bibinfo{author}{\bibfnamefont{F.}~\bibnamefont{Schedin}},
  \bibinfo{author}{\bibfnamefont{M.~I.} \bibnamefont{Katsnelson}},
  \bibinfo{author}{\bibfnamefont{R.}~\bibnamefont{Yang}},
  \bibinfo{author}{\bibfnamefont{E.~H.} \bibnamefont{Hill}},
  \bibinfo{author}{\bibfnamefont{K.~S.} \bibnamefont{Novoselov}},
  \bibnamefont{and} \bibinfo{author}{\bibfnamefont{A.}~\bibnamefont{K.Geim}},
  \bibinfo{journal}{Science} \textbf{\bibinfo{volume}{320}},
  \bibinfo{pages}{356} (\bibinfo{year}{2008}).

\bibitem[{\citenamefont{Brey and Fertig}(2006)}]{BF06}
\bibinfo{author}{\bibfnamefont{L.}~\bibnamefont{Brey}} \bibnamefont{and}
  \bibinfo{author}{\bibfnamefont{H.~A.} \bibnamefont{Fertig}},
  \bibinfo{journal}{Phys. Rev. B} \textbf{\bibinfo{volume}{73}},
  \bibinfo{pages}{235411} (\bibinfo{year}{2006}).

\bibitem[{\citenamefont{Prada et~al.}(2007)\citenamefont{Prada, San-Jos\'e,
  Wunsch, and Guinea}}]{PSWG07}
\bibinfo{author}{\bibfnamefont{E.}~\bibnamefont{Prada}},
  \bibinfo{author}{\bibfnamefont{P.}~\bibnamefont{San-Jos\'e}},
  \bibinfo{author}{\bibfnamefont{B.}~\bibnamefont{Wunsch}}, \bibnamefont{and}
  \bibinfo{author}{\bibfnamefont{F.}~\bibnamefont{Guinea}},
  \bibinfo{journal}{Phys. Rev. B} \textbf{\bibinfo{volume}{75}},
  \bibinfo{pages}{113407} (\bibinfo{year}{2007}).

\bibitem[{\citenamefont{Beenakker and van Houten}(1991)}]{BH91}
\bibinfo{author}{\bibfnamefont{C.~W.~J.} \bibnamefont{Beenakker}}
  \bibnamefont{and} \bibinfo{author}{\bibfnamefont{H.}~\bibnamefont{van
  Houten}}, \bibinfo{journal}{Sol. St. Phys.} \textbf{\bibinfo{volume}{44}},
  \bibinfo{pages}{1} (\bibinfo{year}{1991}).

\bibitem[{\citenamefont{Rycerz and Beenakker}(2007)}]{RB07}
\bibinfo{author}{\bibfnamefont{A.}~\bibnamefont{Rycerz}} \bibnamefont{and}
  \bibinfo{author}{\bibfnamefont{C.~W.~J.} \bibnamefont{Beenakker}}
  (\bibinfo{year}{2007}), \eprint{arXiv:0709.3397}.

\bibitem[{\citenamefont{Rycerk}(2008)}]{R07}
\bibinfo{author}{\bibfnamefont{A.}~\bibnamefont{Rycerk}},
  \bibinfo{journal}{Phys. St. Sol. (a)} \textbf{\bibinfo{volume}{205}},
  \bibinfo{pages}{1281} (\bibinfo{year}{2008}).

\bibitem[{\citenamefont{Iyengar et~al.}(2008)\citenamefont{Iyengar, Luo,
  Fertig, and Brey}}]{ILFB08}
\bibinfo{author}{\bibfnamefont{A.}~\bibnamefont{Iyengar}},
  \bibinfo{author}{\bibfnamefont{T.}~\bibnamefont{Luo}},
  \bibinfo{author}{\bibfnamefont{H.~A.} \bibnamefont{Fertig}},
  \bibnamefont{and} \bibinfo{author}{\bibfnamefont{L.}~\bibnamefont{Brey}}
  (\bibinfo{year}{2008}), \eprint{arXiv:0804.0246}.

\bibitem[{\citenamefont{Akhmerov et~al.}(2007)\citenamefont{Akhmerov,
  Bardarson, Rycerz, and Beenakker}}]{ABRB07}
\bibinfo{author}{\bibfnamefont{A.~R.} \bibnamefont{Akhmerov}},
  \bibinfo{author}{\bibfnamefont{J.~H.} \bibnamefont{Bardarson}},
  \bibinfo{author}{\bibfnamefont{A.}~\bibnamefont{Rycerz}}, \bibnamefont{and}
  \bibinfo{author}{\bibfnamefont{C.~W.~J.} \bibnamefont{Beenakker}},
  \bibinfo{journal}{Phys. Rev. B} \textbf{\bibinfo{volume}{77}},
  \bibinfo{pages}{205416} (\bibinfo{year}{2007}).

\bibitem[{\citenamefont{Abanin and Levitov}(2008)}]{AL08}
\bibinfo{author}{\bibfnamefont{D.~A.} \bibnamefont{Abanin}} \bibnamefont{and}
  \bibinfo{author}{\bibfnamefont{L.~S.} \bibnamefont{Levitov}},
  \bibinfo{journal}{Phys. Rev. B} \textbf{\bibinfo{volume}{78}},
  \bibinfo{pages}{035416} (\bibinfo{year}{2008}).

\end{thebibliography}
\end{document}